\documentclass[12pt]{article}
\usepackage{epsfig, amsmath, amssymb}
\usepackage{graphicx,psfrag} 
\newcommand{\be}{\begin{equation}}
\newcommand{\ee}{\end{equation}}
\newcommand{\bea}{\begin{eqnarray}}
\newcommand{\eea}{\end{eqnarray}}
\newcommand{\bA}{\begin{array}}
\newcommand{\eA}{\end{array}}
\newcommand{\bc}{\begin{center}}
\newcommand{\ec}{\end{center}}
\newcommand{\al}{\alpha}

\newcommand{\ra}{\rightarrow}
\newcommand{\del}{\partial}

\newcommand{\ie}{{\it i.e.}}
\newcommand{\eg}{{\it e.g.}}

\begin{document}


\begin{titlepage}

\bc

\hfill 
\\         [40mm]

{\Huge On a lightlike limit of entanglement} 
\vspace{16mm}

{\large K.~Narayan} \\
\vspace{3mm}
{\small \it Chennai Mathematical Institute, \\}
{\small \it SIPCOT IT Park, Siruseri 603103, India.\\}

\ec
\medskip
\vspace{40mm}

\begin{abstract}
We study certain classes of $g_{++}$ deformations of theories arising
in gauge/string realizations of nonrelativistic holography, some of
which pertain to $z=2$ Lifshitz theories while others (pertaining to
hyperscaling violation) comprise certain classes of excited states.
Building on previous work, we consider holographic entanglement
entropy for spacelike strip subsystems in a highly boosted (lightlike)
limit, where the strip is stretched along the null $x^+$-plane. The
leading divergence in entanglement in this null limit for these
excited states is milder than the usual area law for spacelike
subsystems in ground states. For ground states, the entanglement
vanishes, perhaps consistent with ultralocality. We discuss this
briefly from a field theory perspective. We also present some simple
free lightfront field theory examples in excited states where
correlators are nonvanishing.
\end{abstract}

\end{titlepage}



\section{Introduction}

Certain gauge/string realizations in nonrelativistic generalizations 
of gauge/gravity duality \cite{AdSCFT} involve spacetimes of the form
\be\label{AdSnull}
ds^2 = {R^2\over r^2} (-2dx^+dx^-+ dx_i^2 + dr^2) + 
R^2g_{++} (dx^+)^2 + R^2d\Omega_S^2\ .
\ee 
The compact space $d\Omega_S^2$ arising in string/M-theory 
realizations will not play much role in what follows.
For $g_{++}=0$, we have $AdS$ in lightcone coordinates. When
$g_{++}>0$, dimensional reduction along the bulk $x^+$-direction 
(regarded compact) gives interesting theories in lower 
dimensions, with time $t\equiv x^-$.
Explicit examples include $z=2$ Lifshitz in bulk $d$-dim
\cite{Kachru:2008yh,Taylor:2008tg} arising from $x^+$-reduction (and 
on $S$) of non-normalizable null deformations of $AdS_{d+1}\times S$ 
\cite{Balasubramanian:2010uk,Donos:2010tu} (also \cite{Gregory:2010gx,
Donos:2010ax,Cassani:2011sv,Halmagyi:2011xh,Chemissany:2011mb,
Narayan:2011az,Costa:2010cn}),
\be\label{adsLif}
ds^2 = {R^2\over r^2} [-2dx^+dx^- + dx_i^2 + dr^2] + K^2 R^2 (dx^+)^2 
\ \ \longrightarrow\ \ \
ds^2=-{dt^2\over r^4}+{\sum_{i=1}^{d_i}dx_i^2+dr^2\over r^2}\ .
\ee
These exhibit Lifshitz scaling\ 
$t\ra \lambda^z t ,\ \ x_i\ra \lambda x_i ,\ \ r\ra \lambda r$, 
with dynamical exponent $z$, the lower dimensional theory 
arising in gravity theories with negative cosmological 
constant and massive abelian gauge fields. Other examples include 
$AdS_{d+1}$ plane waves \cite{Narayan:2012hk,Singh:2012un,Narayan:2012wn} 
which give rise to hyperscaling violation,
\bea\label{adspw}
ds^2 &=& {R^2\over r^2} [-2dx^+dx^- + dx_i^2 + dr^2] + R^2Qr^{d-2} (dx^+)^2 
\quad \longrightarrow\\
ds^2 &=& r^{2 \theta \over d_i} \Big(-{dt^2 \over r^{2z}} + 
{\sum_{i=1}^{d_i}  dx_i^2 + dr^2 \over r^2 }\Big), \quad\ \ 
z={d-2\over 2}+2 ,\quad \theta={d-2\over 2}\ ,\quad d_i=d-2 .\ \ 
\label{hypviol}
\eea
The normalizable $g_{++}$ deformation here represents a state in 
the dual CFT with uniform energy-momentum density flux $T_{++}\sim Q$. 
Upon $x^+$-reduction, we obtain a hyperscaling violating
(or conformally Lifshitz) background with Lifshitz $z$ and
hyperscaling violating $\theta$ exponents above, and $d_i$ the
boundary spatial dimension \cite{Gubser:2009qt,Cadoni:2009xm,
Goldstein:2009cv,Charmousis:2010zz,Ogawa:2011bz,Huijse:2011ef,Dong:2012se,
Dey:2012tg} (and many others in the now large literature). These 
hyperscaling violating theories arise in Einstein-Maxwell-scalar
theories (which arise from the higher dimensional description 
under $x^+$-reduction). Some of these exhibit interesting 
entanglement behaviour \cite{Ogawa:2011bz,Huijse:2011ef,Dong:2012se}.

The $x^+$-dimensional reduction implies that the lower dimensional
theory has a time coordinate $t$ identified with lightcone time $x^-$
in the higher dimensional theory (upstairs). It is thus interesting 
to isolate a lightlike limit upstairs (in particular of
entanglement entropy) which describes corresponding features of the
lower dimensional theory.  In the Ryu-Takayanagi holographic
entanglement description \cite{Ryu:2006bv,Ryu:2006ef,HRT}, this
involves calculating the area of an extremal surface on a constant
$x^-$ slice (which is spacelike in the bulk since
$g^{--}<0$). Building on the study of entanglement entropy in $AdS$
plane waves \cite{Narayan:2012ks}, we consider a highly boosted limit
of spacelike strip subsystems: in this lightlike limit, the strip is
stretched along the null $x^+$-plane\ (this is somewhat different 
from the discussions in \cite{Bousso:2014sda,
Bousso:2014uxa}).  This arises in a certain regime in the dual 
field theory upstairs involving the UV cutoff and the energy scale
contained in $g_{++}$ (sec.~2). The entanglement entropy is
calculable exactly in the large $N$ gravity approximation, the
entanglement integrals mapping to those in an auxiliary $AdS$
space. Similar structures arise in boosted black branes and
nonconformal brane plane waves \cite{Narayan:2013qga}. The finite
cutoff-independent part in the Lifshitz case is larger than that for
$AdS$, perhaps reflecting the IR behaviour. The leading divergence in
this lightlike regime for these excited states dual to $AdS$ plane
waves is milder than the usual area law divergence for spacelike
subsystems in ground states \cite{AreaLaw}. We give some comments from
a field theory perspective (sec.~3): ultralocality \cite{Wall:2011hj}
suggests vanishing entanglement for the ground state. We present some
simple calculations in free lightfront quantization of nonvanishing 
correlation functions in excited states, suggesting nonvanishing 
entanglement at weak coupling. Sec.~4 has some discussion.

\section{Entanglement entropy in a null limit}

We want to consider a lightlike limit of holographic entanglement 
entropy for the spacetimes (\ref{AdSnull}), building on that for 
the $AdS_5$ plane wave in \cite{Narayan:2012ks}.\ 
Consider a strip-like subsystem with width $\Delta x = l$ along some 
$x\in\{x_i\}$ (the others labelled $y_i$), and extended along the 
$x^+$-direction, 
\be\label{subsysx}
x^+=\al\chi,\quad x^-=-\beta\chi,\qquad -{l\over 2}<x\leq {l\over 2},
\qquad -\infty <\chi, y_i<\infty\ ,
\ee
\ie\ wrapping the $y_i, \chi$ directions completely. For 
(\ref{subsysx}), (\ref{AdSnull}), the holographic entanglement is
\be\label{EEg++}
S\ =\ {V_{d-2} R^{d-1}\over 4G_{d+1}} \int {dr\over r^{d-1}} 
 \sqrt{2\al\beta+\al^2g_{++}r^2}\ \sqrt{1+(\del_rx)^2}\  \quad 
\Longrightarrow
\ee
\be 
{l\over 2} = \int_\epsilon^{r_*}  {dr\  Ar^{d-1}\over
\sqrt{2\al\beta + \al^2 g_{++} r^2 - A^2 r^{2d-2}}} ,  \quad\
S = {V_{d-2} R^{d-1}\over 4G_{d+1}} \int_\epsilon^{r_*} 
{dr\over r^{d-1}} {2\al\beta + \al^2 g_{++} r^2 \over 
\sqrt{2\al\beta + \al^2 g_{++} r^2 - A^2 r^{2d-2}}} , \nonumber
\ee
the second line obtained by extremization. $\epsilon$ is the UV cutoff, 
$V_{d-2}=\int (\prod_{i=1}^{d-3} dy_i)d\chi$, and $r_*$ the extremal 
surface turning point where ${dr\over dx}|_{r_*}=0$. We have 
${R^{d-1}\over G_{d+1}}\sim N^2,\ N^{3/2},\ N^3$ for D3- ($AdS_5$), 
M2- ($AdS_4$) and M5-branes ($AdS_7$) respectively.
The usual entanglement entropy corresponds to spacelike strips\ 
$\al=\beta=1$ on a time slice $T={x^++x^-\over\sqrt{2}}=0$ 
and stretched along $x^3={x^+-x^-\over\sqrt{2}}$,\ 
with the familiar area law divergence ${V_{d-2}\over\epsilon^{d-2}}$ 
and subleading terms. The parameters $\al, \beta$, define 
the orientation of the (infinitely extended) strip in the 
$x^\pm$- (or $t,x_3-$) plane. While the entanglement entropy varies 
with $\al, \beta$, the leading divergence is the usual area law 
if the strip is spacelike; likewise finite cutoff-independent pieces 
can be estimated. For instance, the $AdS_5$ plane wave ($g_{++}=Qr^2$,
with $Q$ the holographic energy momentum flux $T_{++}$) gives the
leading divergence $S^{div}\sim N^2{V_2\over\epsilon^2}$~, 
the finite part being\ $S^{fin}\sim N^2V_2\sqrt{Q}\log (lQ^{1/4})$\ 
\cite{Narayan:2012ks}\ (resembling that for a Fermi surface 
with Fermi momentum $k_F\equiv Q^{1/4}$;\ this was the original 
motivation for \cite{Narayan:2012hk}).

The lightlike limit corresponding to null time $x^-$ slices, is 
obtained with\ $\al=1, \beta=0$, 
\be\label{EEnull}
l = \Delta x = 2\int_\epsilon^{r_*} {dr\ Ar^{d-1}\over
\sqrt{g_{++} r^2 - A^2 r^{2d-2}}}\ , \qquad\
S = {V_{d-2} R^{d-1}\over 4G_{d+1}} \int_\epsilon^{r_*} 
{dr\over r^{d-1}} {g_{++} r^2 \over \sqrt{g_{++} r^2 - A^2 r^{2d-2}}}\ .
\ee
This is natural from the point of view of the lower dimensional 
theory obtained by $x^+$-reduction, where $x^-$ becomes time $t$ below.
It can be defined physically as a highly boosted limit of a spacelike 
subsystem. Consider the boost $x^\pm\ra \lambda^{\pm 1} x^\pm$: this 
transforms the subsystem (\ref{subsysx}) with $\al=1=\beta$ to 
$\al=\lambda\gg 1,\ \beta={1\over\lambda}\ll 1$, giving
\be\label{EEboostx}
{l\over 2} = \int_\epsilon^{r_*} {dr\ Ar^{d-1}\over
\sqrt{2 + \lambda^2 g_{++} r^2 - A^2 r^{2d-2}}} ,  \quad\ \
S = {V_{d-2} R^{d-1}\over 4G_{d+1}} \int_\epsilon^{r_*} 
{dr\over r^{d-1}} {2 + \lambda^2 g_{++} r^2 \over 
\sqrt{2 + \lambda^2 g_{++} r^2 - A^2 r^{2d-2}}} .
\ee
Under the boost transformation, we see that the entangling surface 
(correspondingly the induced metric and its area) changes: the 
entanglement entropy is not boost invariant. For $\lambda=1$, as 
$r\ra r_{min}=\epsilon$, the term $\lambda^2 g_{++}r^2$ in 
(\ref{EEboostx}) becomes small compared with the leading `2'.
As the boost $\lambda$ increases, 
$\lambda^2 g_{++}r^2|_{min}=\lambda^2 g_{++}r^2|_{r=\epsilon}$ 
increases: thus $\lambda^2 g_{++}r^2$ increases for all $r$ in 
(\ref{EEboostx}) since $r>r_{min}=\epsilon$. Thus in the regime
\be\label{EEnullregime}
\lambda^2 g_{++}(\epsilon) \epsilon^2\ \gtrsim\ 1\ ,
\ee
the `2' can be dropped compared with $\lambda^2g_{++}r^2$: thus 
(\ref{EEboostx}) is well approximated by the null case 
(\ref{EEnull}).\ Here $g_{++}$ representing either excited 
states or a CFT-deformation contains a new length scale and
(\ref{EEnullregime}) gives a new regime where the $g_{++}$ length
scale (after boost) is comparable to the ultraviolet cutoff
$\epsilon$.  Heuristically, the boundary $r=\epsilon$ now dips 
sufficiently inward in the bulk to feel the departure from pure
$AdS$ due to $g_{++}$. The strip subsystem (\ref{subsysx}) now 
parametrized as\ 
$x^-=0,\ -{l\over 2}<x\leq {l\over 2},\ -\infty <x^+, y_i<\infty$,\ 
is stretched along the null $x^+$-plane. The expressions 
(\ref{EEboostx}) can be analysed further in this regime (with 
specific examples later). The turning point of the surface is the 
location where the denominator in (\ref{EEboostx}) vanishes, \ie\ 
$2 + \lambda^2 g_{++}(r_*) r_*^2 - A^2 r_*^{2d-2} = 0$. In the 
lightlike regime, this is approximated as\ $\lambda^2 g_{++}(r_*) 
\sim A^2 r_*^{2d-4}$\ (\ie\ where the denominator 
in (\ref{EEnull}) vanishes). The parameter $A$ thus scales as 
$A^2\sim \lambda^2 g_{++}(r_*)r_*^{4-2d}$. The width integral becomes\
$l \sim \int_0^{r_*} {dr (A/\sqrt{\lambda^2g_{++}}) r^{d-2}\over
\sqrt{1-(A^2/(\lambda^2 g_{++}))r^{2d-4}}}$ showing $l\sim r_*$. 
The entanglement becomes\ 
$S = {V_{d-2} R^{d-1}\over 4G_{d+1}} \int_\epsilon^{r_*} 
{dr\over r^{d-1}} {\sqrt{\lambda^2 g_{++} r^2} \over 
\sqrt{1 - A^2 r^{2d-2}/(\lambda^2 g_{++} r^2)}}$, showing the leading 
divergence\ 
$\int_\epsilon dr {\sqrt{\lambda^2 g_{++} r^2}\over r^{d-1}}\sim 
{\sqrt{\lambda^2g_{++}(\epsilon)}\over \epsilon^{d-3}}$.\
This new entanglement behaviour in this null limit of the higher 
dimensional theory is equivalent to (and essentially implied by) 
the entanglement behaviour in the lower dimensional descriptions via 
the gauge/string realizations (\ref{AdSnull}), and in particular to 
that in hyperscaling violating backgrounds 
\cite{Ogawa:2011bz,Huijse:2011ef,Dong:2012se} in the gauge/string 
realizations (\ref{adspw}) (\ref{hypviol}). We will see 
this in greater detail below.

\subsection{$AdS$ null deformations,\ $g_{++}=K^2$\ \ $\ra$\ \  $z=2$ Lifshitz}

Consider first the $AdS_{d+1}$ null deformations (\ref{adsLif})
which give $z=2$ Lifshitz in bulk $d$-dim upon $x^+$-reduction 
\cite{Balasubramanian:2010uk,Donos:2010tu}, with $x^-\equiv t$.
The $g_{++}$ mode independent of $r$ is sourced by other matter: 
\eg\ an axion source $c_0=Kx^+$ gives $g_{++}\sim(\del_+c_0)^2=K^2$, 
with $K$ a constant of mass dimension one 
\cite{Narayan:2011az,Chemissany:2011mb,Balasubramanian:2011ua}.
The SL(2,Z) duality $\tau\ra\tau+1$, \ie\ $c_0\ra c_0+1$, of IIB 
string theory means the axion profile is effectively $x^+$-periodic 
with periodicity ${1\over K}$~. For noncompact $x^+$, the 
$x^-,x_i$-subspace has $z=2$ Lifshitz scaling, which in the bulk is\ 
$(x^-,x_i,r)\ra (\lambda^2x^-,\lambda x_i,\lambda r)$.

In the upstairs description, we consider holographic 
entanglement entropy for the $AdS$-Lifshitz deformation: structurally 
this is similar to \cite{Narayan:2012ks} for $AdS$ plane waves. 
For the strip subsystem (\ref{subsysx}), the entanglement entropy 
is (\ref{EEg++}) with $g_{++}=K^2$.
The spacelike subsystem (\ref{subsysx}) has $\al=1=\beta$.
With $\epsilon\ll {1\over K}$ the leading divergence is the area law\ 
$\sim {V_{d-2}\over\epsilon^{d-2}}$ in boundary $d$-dim, \ie\ in 
$AdS_{d+1}$. The turning point is\ 
$2 + K^2 r_*^2 - A^2 r_*^{2d-2} = 0$. For large width $l$ and large $K$, 
this is\ $K^2 \sim A^2 r_*^{2d-4}$, giving\ $l\sim r_*$\ 
($\sim\int_\epsilon^{r_*} {dr\ (A/K) r^{d-2}\over\sqrt{1-(A^2/K^2)r^{2d-4}}}$).

The subsystem (\ref{subsysx}) with $\al=1=\beta$ under the boost 
$x^\pm\ra \lambda^{\pm 1} x^\pm$, with large $\lambda$, leads to 
the regime (\ref{EEnullregime}) governing the lightlike limit. In 
the unboosted theory, this is the regime
\be\label{lifregime}
K^2 \epsilon^2\ \gtrsim\ 1\ ,\qquad\ \ie\ \qquad 
K\ \gtrsim\ \epsilon^{-1}\ ,
\ee
\ie\ the Lifshitz-deformation scale $K$ is comparable to the UV scale 
$\epsilon^{-1}$.\
Equivalently, we have redefined $\lambda K\ra K$ in (\ref{EEnullregime}).
Since $r>r_{min}=\epsilon$, the entanglement becomes  
\be\label{lifEEnull}
S \sim\  {V_{d-2} R^{d-1}\over 4G_{d+1}} \int_\epsilon^{r_*} 
{dr\over r^{d-2}} { K \over\sqrt{1 - (A^2/K^2) r^{2d-4}}}\
\sim\ {R^{d-1} \over 4G_{d+1}}\ {V_{d-2} K\over d-3} 
\Big({1\over\epsilon^{d-3}}-c_d{1\over l^{d-3}}\Big)\ .
\ee
The expression (\ref{lifEEnull}), identical to that in some auxiliary 
$AdS_d$ space, is a null limit, $\al\sim 1, \beta\sim 0$, 
(\ref{EEnull}) of (\ref{EEboostx}). The $AdS_4$ deformed theory 
gives\ $V_1KN^{3/2} \log{l\over\epsilon}$.

This resembles the entanglement in the lower dimensional Lifshitz$_d$
theory\footnote{From (\ref{adsLif}), $x^+$-reduction gives\ 
$ds_d^2= (KR)^{2/(d-2)} R^2 (-{(dx^-)^2\over K^2r^4} + 
{dx^2+dy^2+dr^2\over r^2})$ retaining the length scales of the 
$AdS_{d+1}$ theory. The entanglement entropy of this Lif$_d$ theory 
for a strip with width $l$ along $x$ is the area of the minimal 
surface on a $x^-\equiv t$ slice,\ 
$S \sim {K R^{d-1}\over 4G_d} V_{d-3} 
({1\over\epsilon^{d-3}} - {1\over l^{d-3}})$.\ Using 
$V_{d-2}=V_{d-3} V_+$ with $V_+={1\over K}$, and 
${V_+\over G_{d+1}}={1\over G_d}$, this is equivalent to 
(\ref{lifEEnull}).}: 
\ie\ in the regime (\ref{EEnullregime}), (\ref{lifregime}), on
length scales longer than the axion variation scale ${1\over K}$ 
(with $\epsilon\gtrsim {1\over K}$),  entanglement in
the higher dimensional $AdS_{d+1}$ deformed theory resembles that in
the Lifshitz$_d$ theory, with leading divergence the area law
${1\over\epsilon^{d-3}}$ in boundary $d-1$ dimensions.  For length
scales much shorter than ${1\over K}$, the Lifshitz deformation has
not turned on so the UV structure for $\epsilon\ll {1\over K}$ must 
be as in $AdS_{d+1}$: this reflects in the leading divergence
${V_{d-2}\over \epsilon^{d-2}}$, \ie\ the boundary $d$-dim area law, 
consistent with the Lif$_d$ theory arising only on scales large 
relative to ${1\over K}$. 

The finite entanglement\
$S^{finite}\ \sim\ {R^{d-1}/G_{d+1}\over 3-d} {V_{d-2} K\over l^{d-3}}$\ 
(and $N^{3/2}V_1K\log (lK)$ for $d=3$) in a sense contains the IR
degrees of freedom encoding entanglement: as the width $l$ increases,
the extremal surface dips deeper into the interior ($r_*\sim l$). 
Comparing with $AdS_{d+1}$, the finite part in the Lifshitz case is
larger for $l\gg {1\over K}$.  Perhaps this is a reflection of more
soft modes in the Lifshitz theory (compared with a relativistic CFT)
responsible for the IR singularities 
\cite{Kachru:2008yh,Horowitz:2011gh}, stemming from the $\omega\sim
k^z$ dispersion relation in field theory (see also
\cite{Andrade:2014bsa}).

\subsection{$AdS_{d+1}$ plane waves:\ \ $g_{++}=Qr^{d-2}$}

Now let us consider $AdS_{d+1}$ plane waves (\ref{adspw}). Using 
(\ref{EEboostx}), 
we have the turning point equation\ $2+\lambda^2Qr_*^d-A^2r_*^{2d-2}=0$,
which for  $\lambda^2Qr_*^d \gg 1$ is well approximated by 
$\lambda^2Qr_*^d - A^2 r_*^{2d-2} = 0$. Here the regime 
(\ref{EEnullregime}) is
\be
\lambda^2 Q \epsilon^d \gtrsim 1\ ,\qquad\ \ie \qquad\ 
P_+' \gtrsim\ \epsilon^{-1}\ ,
\ee
\ie\ the elemental lightcone momentum 
$P_+=T_{++}\Delta x^+\Delta^{d-2}x|_\epsilon$ 
at scale $\epsilon$ in the boosted frame is comparable to the UV 
cutoff $\epsilon^{-1}$\ (the transverse area $\Delta^{d-2}x$ is boost 
invariant).\  In this regime, $\lambda^2Qr^d\gtrsim 1$ for 
all $r$-values appearing in (\ref{EEboostx}) since $r>\epsilon$:
the entanglement (\ref{EEboostx}) for the $AdS_{d+1}$ plane wave is 
well-approximated by the corresponding expressions (\ref{EEnull}) at 
null slicing
\be\label{EEboostxAdSpwNull}
{l\over 2} \sim \int_\epsilon^{r_*} dr {(A/\sqrt{\lambda^2 Q})\ 
r^{d/2-1}\over\sqrt{1 - (A^2/\lambda^2 Q) r^{d-2}}} \sim r_* , \quad
S  \sim  {V_{d-2} R^{d-1}\over 4G_{d+1}}
\int_\epsilon^{r_*} {dr\over r^{d/2-1}} { \sqrt{\lambda^2 Q} \over
\sqrt{1 - (A^2/\lambda^2 Q) r^{d-2}}} .
\ee
In this highly boosted limit, we are effectively probing entanglement 
at scales $P_+'\gtrsim \epsilon$.\ Equivalently, the regime 
(\ref{EEnullregime}) moves us into the hyperscaling violating regime. 

From (\ref{EEboostxAdSpwNull}) we see that the null
entanglement integral for the $AdS_{d+1}$ plane wave excited states 
is identical to
that for a spacelike strip subsystem in an auxiliary pure $AdS$ space 
in $({d\over 2}+1)$-dim\ 
(\eg\ using (\ref{EEg++}) with $\al=\beta=1,\ g_{++}=0$): it can be 
evaluated exactly as for $AdS$
\be\label{nullEEresult}
l\sim r_*\ ,\qquad\quad  S \sim {R^{d-1}\over 4G_{d+1}} 
{V_{d-2}\sqrt{\lambda^2 Q}\over d-4}
\Big({1\over\epsilon^{{d\over 2}-2}} - c_d {1\over l^{{d\over 2}-2}}\Big)\ ,
\ee
using the turning point\ $\lambda^2Qr_*^d = A^2 r_*^{2d-2}$, with $c_d$ 
a constant. The lightlike limit of entanglement (\ref{EEboostxAdSpwNull}),
(\ref{nullEEresult}) in the $AdS_{d+1}$ plane wave in the regime 
(\ref{EEnullregime}) is essentially the entanglement with ordinary 
time slicing in a theory in ${d\over 2}$-dim:\ this is in fact the 
lower dimensional theory with hyperscaling violation (after 
$x^+$-reduction), living in an effective dimension\ 
$d_{eff}=d-1-\theta={d\over 2}$, using (\ref{hypviol}). Since
$d_{eff}<d-1$ for $d>2$, the leading divergence is milder than 
the usual area law (see (\ref{EEg++})).
For the $AdS_5$ plane wave dual to 4d SYM CFT excited 
states, (\ref{EEboostxAdSpwNull}) reduces to
\be\label{eeadspwLog}
{l\over 2} = \int_0^{r_*}  
{dr\ r (A/\sqrt{\lambda^2 Q}) \over
\sqrt{1 - {A^2\over\lambda^2 Q} r^2}} \sim (\#) r_* , \quad
S \sim {V_2 R^3\over 4G_5} \int_\epsilon^{r_*}  
{dr\ \sqrt{\lambda^2 Q} \over r \sqrt{1 - {A^2\over \lambda^2 Q} r^2}} \sim
N^2 V_2 \sqrt{\lambda^2 Q}\ \log {l\over\epsilon}  ,
\ee
discussed previously in \cite{Narayan:2012wn,Narayan:2012ks}.
This is reminiscent of a 2-dim CFT with central charge $N^2V_2\sqrt{Q}$.
From (\ref{EEboostxAdSpwNull}), the $AdS_4$ plane wave (with just 
one transverse dimension) gives\
$S \sim\ V_1 N^{3/2} \sqrt{\lambda^2 Q} (\sqrt{l}-\sqrt{\epsilon})$,\
with no divergence, valid if $\epsilon\gtrsim {1\over (\lambda^2 Q)^{1/3}}$.
(The change in entanglement $\Delta S$ is as in the lower dimensional 
theory.)
Pure $AdS$ corresponds to the CFT ground state: for null
slicing (\ref{EEnull}), entanglement entropy vanishes since $g_{++}=0$.
From (\ref{AdSnull}), we see the entangling surface degenerates.

\subsection{Boosted black branes}

Consider black D3-branes\ 
$ds^2={R^2\over r^2} [-(1-r_0^4r^4) dt^2 + dx_3^2 +\sum_{i=1}^2 dx_i^2]
+ {R^2dr^2\over r^2 (1-r_0^4r^4)}$ and lightcone coordinates $x^\pm$ as 
$T={x^++x^-\over\sqrt{2}},\ x^3={x^+-x^-\over\sqrt{2}}$.
Boosting as\
$x^\pm\ra \lambda^{\pm 1}x^\pm$ \cite{Maldacena:2008wh} gives\
\bc
$ds^2 = {R^2\over r^2} \big( -2dx^+dx^-+ {r_0^4 r^4\over 2}
(\lambda dx^++\lambda^{-1} dx^-)^2 +\sum_{i=1}^2 dx_i^2\big)
+ {R^2 dr^2\over r^2 (1-r_0^4r^4)}$~. \ec
For large boost $\lambda$ and low temperature $r_0$, with 
$Q={\lambda^2r_0^4\over 2}$, this approaches the (near extremal) 
$AdS_5$ plane wave (\ref{adspw}). For the subsystem (\ref{subsysx}), 
we have the entanglement area functional \cite{Narayan:2012ks}\ 
$S = {V_2R^3\over 4G_5} \int {dr\over r^3} 
\sqrt{(\del_rx)^2+{1\over 1-r_0^4r^4}}
\sqrt{2\al\beta + {\lambda^2r_0^4r^4\over 2} (\al-{\beta\over\lambda^2})^2}$\ 
for the boosted black brane giving
\be\label{EEboostedBB}
S = {V_2R^3\over 4G_5} \int_\epsilon^{r_*} {dr\over r^3} 
{2\al\beta + {\lambda^2r_0^4r^4\over 2} 
(\al-{\beta\over\lambda^2})^2\over \sqrt{1-r_0^4r^4}
\sqrt{2\al\beta + {\lambda^2r_0^4r^4\over 2} (\al-{\beta\over\lambda^2})^2 
- A^2r^6}}\ ,
\ee
and $\Delta x = l \sim r_*$, after extremizing.
For large boost $\lambda$, we have ${\beta\over\lambda^2}\ra 0$ and 
in the regime\ $\lambda^2r_0^4\epsilon^4 \gtrsim 1$, the second term 
in the radical (containing $\lambda^2r_0^4r^4$) is always greater 
than the first: thus this resembles the entanglement\ 
$S \sim {V_2R^3\over 4G_5} \int_\epsilon^{r_*} {dr\over r} 
{(\lambda^2r_0^4/2) \over \sqrt{1-r_0^4r^4}
\sqrt{(\lambda^2r_0^4/2) - A^2r^2}}$\ for a null subsystem 
(setting $\al\sim 1$).
The boost parameter introduces a separation of scales. 
The entanglement (\ref{EEboostedBB}) in the regime\ 
${1\over\sqrt{\lambda} r_0} \lesssim \epsilon \ll r_* \ll {1\over r_0}$, 
becomes\  $S \sim N^2 V_2\sqrt{Q}\ \log {l\over\epsilon}$,
as for the $AdS_5$ plane wave (\ref{eeadspwLog}) which is regulated
by the boosted black brane.\  For an 
unboosted brane $\lambda=1$, this null entanglement ($\al=1,\beta=0$) 
is\ $S \sim\ N^2 V_2 {r_0^2\over \sqrt{2}}\ \log {l\over\epsilon} + 
{N^2V_2\over \sqrt{2}} {r_0^6\over l^4} + \ldots$\ 
Thus the 4-dim SYM CFT thermal state with null time slicing in the 
regime\ $\epsilon\gtrsim {1\over r_0}$ exhibits entanglement 
(\ref{EEboostedBB}) with a leading logarithmic divergence. 
In the far IR limit $l\sim r_*\sim {1\over r_0}$, this is extensive, 
resembling the usual thermal entropy, expected from the lower 
dimensional theory.

\subsection{Nonconformal D-brane plane waves}

The string metric and dilaton for nonconformal $Dp$-brane plane waves 
\cite{Narayan:2013qga} (see also \cite{Singh:2012un}) are 
\bea\label{Dpnullnorm}
&& ds^2_{st} = {r^{(7-p)/2}\over R_p^{(7-p)/2}} dx_{\parallel}^2 + 
{G_{10} Q_p\over R_p^{(7-p)/2}} {(dx^+)^2\over r^{(7-p)/2}} 
+ R_p^{(7-p)/2} {dr^2\over r^{(7-p)/2}} 
+ R_p^{(7-p)/2} r^{(p-3)/2} d\Omega_{8-p}^2\ ,\nonumber\\
&& e^\Phi = g_s \Big({R_p^{7-p}\over r^{7-p}}\Big)^{{3-p\over 4}} ;
\quad g_{YM}^2\sim g_s {\al'}^{(p-3)/2}\ ,\qquad 
R_p^{7-p}\sim g_{YM}^2N {\al'}^{5-p} \sim g_sN {\al'}^{(7-p)/2}\ .\quad
\eea
(The ultraviolet is towards large $r$ here.)
As in the conformal cases, the $g_{++}$ term is obtained 
from the non-conformal finite temperature solutions 
\cite{Itzhaki:1998dd} in a large boost $\lambda$, low temperature 
limit holding the energy-momentum density\ 
${\lambda^2\varepsilon_{p+1}\over 2}\equiv Q_p$ fixed\ (with 
$(U_0\al')^{7-p}\sim G_{10} \varepsilon_{p+1}$ and $U={r\over\al'}$).
These describe strongly coupled Yang-Mills theories with  
energy flux $T_{++}$.
The Einstein metric $ds_E^2 = e^{-\Phi/2} ds_{st}^2$, 
upon dimensional reduction on $S^{8-p}$ and the $x^+$-direction, 
gives hyperscaling violating spacetimes (\ref{hypviol}) with 
exponents\   
$\theta={p^2-6p+7\over p-5} ,\ z={2(p-6)\over p-5}$.\ 
With $r_{UV}$ the UV cutoff and $r_*$ the turning point, the 
entanglement entropy \cite{Narayan:2013qga} for the subsystem 
(\ref{subsysx}) is\
$S_A 
 \sim {V_{p-1} R_p^{7-p}\over G_{10}} \int^{r_{UV}}_{r_*} dr r\ 
\sqrt{2\beta + \al {G_{10}Q\over r^{7-p}}\ } 
\sqrt{1+{r^{7-p}\over R_p^{7-p}} (\del_rx)^2\ }$.
The lightlike limit $\beta=0$ gives
\be\label{EEspacelike}
l = \Delta x = \int^{r_{UV}}_{r_*} {dr\ A R_p^{7-p}\over r^{8-p} 
\sqrt{{G_{10}Q\over r^{7-p}} - {A^2 R_p^{7-p}\over r^{9-p}} }} ,\qquad 
S \sim {V_{p-1} R_p^{7-p}\over G_{10}} \int^{r_{UV}}_{r_*} dr r\ 
{ {G_{10}Q\over r^{7-p}} \over
\sqrt{{G_{10}Q\over r^{7-p}} - {A^2 R_p^{7-p}\over r^{9-p}} }} ,
\ee
with turning point $G_{10}Q r_*^2 = A^2 R_p^{7-p}$.
These again are similar to spacelike subsystems in pure $AdS$, so that
\be\label{nullEEnonconf}
S \sim\ {V_{p-1}\sqrt{Q} \over (3-p) \sqrt{G_{10}}} 
{R_p^{7-p}\over r^{(3-p)/2}}\Big|^{r_{UV}}_{r_*}\ \sim\ 
{V_{p-1}\sqrt{QN^2}\over 3-p}\
\Big({(\sqrt{g_{YM}^2N})^{d_{eff}-2}\over\epsilon^{d_{eff}-2}}
- {(\sqrt{g_{YM}^2N})^{d_{eff}-2}\over l^{d_{eff}-2}}\Big) .
\ee
We have used (\ref{Dpnullnorm}) to recast the bulk result in 
field theory variables\ $l\sim {R_p^{(7-p)/2}\over r_*^{(5-p)/2}} ,\ 
\epsilon\sim {R_p^{(7-p)/2}\over r_{UV}^{(5-p)/2}}$.\ The effective 
dimension here, after $x^+$-reduction, works out to\ 
$d_{eff}={7-p\over 5-p}=p-\theta$, with $d_{eff}<p$ for the cases 
$p=2,3,4$ of relevance. For ground states $Q=0$, the entanglement 
vanishes.

In terms of the scale-dependent number of degrees of freedom\ 
$N_{eff}(s) = N^2 \big({g_{YM}^2N\over s^{p-3}}\big)^{{p-3\over 5-p}}$ 
\cite{Barbon:2008ut} the finite part of entanglement can be written 
\cite{Narayan:2013qga} as\ 
${\sqrt{N_{eff}(l)}\over 3-p}\ {V_{p-1}\sqrt{Q} \over l^{(p-3)/2}}$~:\
the form (\ref{nullEEnonconf}) makes manifest the lightlike limit as 
equivalent to the lower dimensional theory.

\section{On null entanglement in field theory excited states}

We have discussed entanglement entropy in the strongly coupled regime 
for strip subsystems (\ref{subsysx}) in a lightlike limit using the 
Ryu-Takayanagi prescription. We will now make some comments on field 
theory in excited states with energy-momentum density $T_{++}=Q$ 
nonzero.  We recall (\ref{nullEEresult}) for CFT$_d$ excited states,
\ie\
\be\label{nullEEcft}
S_{div}\ \sim\ N^2 \sqrt{Q}\ {V_{d-2}\over\epsilon^{d_{eff}-2}}\
=\ N^2 \sqrt{Q\epsilon^d}\ {V_{d-2}\over\epsilon^{d-2}}\ , 
\qquad\qquad\  d_{eff}={d\over 2}\ .
\ee
The leading divergence (\ref{nullEEcft}) is less severe than the 
usual area law ${V_{d-2}\over\epsilon^{d-2}}$ for the ground state 
\cite{AreaLaw}. For $d=4$, we obtain a logarithmic divergence, the
entanglement scaling as\ $\log {l\over\epsilon}$.  Heuristicically,
the short distance ``entangling degrees of freedom'' or ``partons''
with this null slicing are fewer than with ordinary time slicing. For
the 3d Chern-Simons CFTs arising on M2-branes,\ we replace $N^2\ra
N^{3/2}$: there is no divergence in this case.

For the ground state, (\ref{nullEEcft}) vanishes, suggesting that
there are no partons encoding entanglement in the null limit: indeed
the $\sqrt{Q\epsilon^d}$ factor reflects energy enhancement. This
appears consistent with the ultralocality axiom discussed in
\cite{Wall:2011hj}, \ie\ all $n$-point functions of field operators on
the constant $x^-$ null surface located on $n$ distinct generators,
equivalently at points with spacelike separation $\Delta x_i>0$,
vanish (see also \cite{Bousso:2014sda,Bousso:2014uxa}).

By ultralocality, correlation functions vanish and so entanglement 
must also vanish. While this is true for ground states, it may not 
hold for excited states, consistent with the entanglement exhibiting 
the milder divergence we have discussed so far. In this regard, we 
now heuristically discuss some simple free field correlation 
functions in lightfront quantization (see \eg\ \cite{Susskind:1994vu,
Harindranath:1996hq,Burkardt:1995ct} and references therein; our 
notation here uses $x^-$ as lightcone time however) which corroborate 
this. Consider a 4-dim massless scalar with mode expansion and 
commutation relations
\bea
&& \phi = \int d^2k_i \int_0^\infty {dk_+ \over \sqrt{(2\pi)^3 2k_+}}\ 
\Big( a_{k_+,k_i}  e^{-ik_+x^+-ik_ix^i-i{k_i^2\over 2k_+} x^-}\ +\ 
a^\dag_{k_+,k_i}  e^{ik_+x^++ik_ix^i+i{k_i^2\over 2k_+} x^-}\Big) ,\nonumber\\
&& \qquad
[a_{k_+,k_i}, a^\dag_{k_+',k_i'}] = 
(2\pi)^3\delta(k_+-k_+')\delta^2(k_i-k_i')\ ,
\qquad\ \  a_{k_+,k_i}|0\rangle = 0\ .
\eea
A positive frequency mode has $k_-={k_i^2\over 2k_+}>0$. This imposes
$k_+\geq 0$ in the sum over modes, and excitations must have positive 
lightcone momentum $k_+$: small $k_+$ modes are high energy 
(we will not worry about zero mode issues here). With 
$T_{++}= (\del_+\phi)^2$, the operator $P_+=\int dx^+d^2x_i T_{++}$ 
measuring lightcone momentum is
\be
P_+ = \int dk_+ d^2k_i\ k_+\ a^\dag_{k_+,k_i} a_{k_+,k_i}
\ee
dropping normal ordering terms.\ A simple excited state contributing 
to nonzero $T_{++}$ and so $P_+$ is
\be\label{k+states}
|k_+\neq 0 \rangle\ =\ a^\dag_{k_+} |0\rangle\ , \qquad 
P_+ |k_+\rangle = k_+ |k_+\rangle\ .
\ee
It can be checked that a 2-point function 
$\langle 0| \del_+\phi(x_1)\ \del_+\phi(x_2) |0\rangle$ on an 
$x^-=const$ surface, with $x_i^1$ and $x_i^2$ distinct, in the vacuum or 
ground state vanishes (as do $n$ point functions): these contain terms 
of the form\ $\int d^2k_i e^{-ik_i\Delta x^i}\sim \delta(\Delta x_i)$ 
which vanish for $\Delta x_i\neq 0$. For excited states of the form 
above, such a 2-point function is
\be
\langle k_+|\ \del_+\phi(x_+^1,x_i^1) \del_+\phi(x_+^2,x_i^2)\ |k_+\rangle\
\sim\ \langle 0| a_{k_+}\ \del_+\phi(x_+^1,x_i^1) \del_+\phi(x_+^2,x_i^2)\ 
a^\dag_{k_+} |0\rangle\ \neq\ 0\ .
\ee
The term\ $\langle 0| \int_{k_1} \int_{k_2} (ik_+^1) (-ik_+^2) \ 
a^{}_{k_+} a^\dag_{k_+^1,k_i^1} a_{k_+^2,k_i^2} a^\dag_{k_+} 
e^{ik^1\cdot x^1-ik^2\cdot x^2} |0\rangle$,\ with\
$\int_k\equiv \int d^2k_i \int_0^\infty {dk_+\over\sqrt{(2\pi)^3 2k_+}}$\ 
and $\delta(k^1,k^2)\equiv \delta(k_+^1-k_+^2)\delta^2(k_i^1-k_i^2)$\ \
gives\ \ $\langle 0| \int_{k_1} \int_{k_2} (ik_+^1) (-ik_+^2) \ 
\delta(k_+,2)\delta(1,k_+) e^{ik^2\cdot x^2 - ik^1\cdot x^1} |0\rangle$ 
$\sim k_+ e^{ik_+\Delta x^+_{12}}$.\ 
Another similar term arises with $1\leftrightarrow 2$. 
These terms are nonzero for generic $x_1,x_2$.
This is quite different from a 4-point function in the ground state,
which also vanishes.  Similar calculations apply for higher point
correlation functions (in states like 
$\prod_i a_{k_{i,+}}^\dag |0\rangle$). Gauge fields in 
lightcone gauge $A_+=0$ have some structural similarities. It would
be interesting to study correlation functions
systematically using smearing functions, constructing normalizable
states (\ref{k+states}) etc: in this case, $\del_+\phi$ gives a
good operator-valued distribution (possible UV divergences in its
correlators, at small $k_+$ or equivalently large $k_-$, can be
controlled by smearing along the lightfront directions).
Entanglement entropy is related to 2-point correlation functions 
using correlation matrices (see \eg\ \cite{peschel}), 
suggesting nonvanishing entanglement entropy at weak coupling: 
this would be interesting to study in detail.

The expression (\ref{nullEEcft}) is recovered if the field theory 
is taken to live on a space
\be\label{YMmet}
ds^2 = -2dx^+dx^- + g^2 (dx^+)^2 + \sum_{i=1}^{d-2} dx_i^2\ ,
\qquad\qquad  g^2 = T_{++}\epsilon^d \gtrsim 1\ .
\ee
The dimensionless parameter $g^2$ is proportional to the energy density 
$T_{++}$ and contains appropriate powers of the cutoff 
$\epsilon$. In the ground state, $T_{++}=0\ \Rightarrow\ g^2=0$ so 
the $x^-$ direction is null, representing lightcone time. For 
excited states with $T_{++}=Q$, the $x^+$ direction formerly null 
``puffs up''.  Discussions of lightfront quantization often use 
the space (\ref{YMmet}) with $g^2$ a regulator (see 
\eg\ \cite{Burkardt:1995ct,Hellerman:1997yu}): in the present case 
$g^2$ is in a sense physical, as reflected in entanglement entropy. 
Now a constant $x^-$ surface is spacelike (with a timelike normal 
since $g^{--}=-g^2<0$).\ The relation between $x^-$ and lightcone 
time $X^-$ is\ $x^- = X^- + {g^2\over 2} x^+$:\ thus $x^-$ is 
not null but timelike in the boundary theory if $g^2\sim O(1)$.\
This corroborates with the bulk: for $AdS$ plane waves (\ref{adspw}), 
the field theory lives on the boundary $r=\epsilon$ with metric 
(\ref{YMmet}). The condition $g^2\sim O(1)$ is equivalent 
to the lightlike regime (\ref{EEnullregime}) with new divergence 
behaviour in the bulk entanglement.

We consider entanglement in this weak coupling field theory on the 
space (\ref{YMmet}) for a strip (\ref{subsysx}) with width 
$l$. When $g^2\ll 1$ and we use time $T={x^++x^-\over\sqrt{2}}$, the 
entanglement leading divergence is the usual area law\
$S_{div}^t\ \sim\ N^2 {V_{x^3} V_{y_i}\over \epsilon^{d-2}}$.
Now with $g^2\sim O(1)$ in the regime (\ref{EEnullregime}), we 
consider $x^-$ as time, with entanglement on constant $x^-$ slices. 
For a strip (\ref{subsysx}) stretched along $x^+$, the usual area 
law divergence for ground states in the space 
(\ref{YMmet}) is 
\be\label{EEg2}
S_{div}\ \sim\ N^2 {V_{x^+} V_{y_i}\over \epsilon^{d-2}} 
\ =\ N^2 V_{d-2} {\sqrt{T_{++}\epsilon^d~}\over \epsilon^{d-2}} 
\ =\ N^2 \sqrt{Q} {V_{d-2}\over \epsilon^{d_{eff}-2}}\ , \qquad\quad 
d_{eff}={d\over 2}\ .
\ee
This is in agreement with the bulk scaling (\ref{nullEEresult}) 
(\ref{nullEEcft}) in the lightlike regime, with 
an effective scaling dimension (in the hyperscaling violating regime) 
arising from the energetic ``puffing up'' of the originally null 
$x^+$ direction. The $g^2$ term in a sense encodes the backreaction 
of the excited state in the original space (with $g^2=0$ 
and $x^+$ null): this gives the usual area law (\ref{EEg2}) for 
the new ground state in the backreacted space (\ref{YMmet}).

For nonconformal theories, the space on which the field theory lives 
is again (\ref{YMmet}), with 
\be\label{YMmetnonconf}
g\ =\ \sqrt{{G_{10}Q\over r_{UV}^{7-p}}}\ =\ 
{\sqrt{Q}\ \epsilon^{d_{eff}} \over N (g_{YM}^2N)^{d_{eff}-2}}\ ,
\qquad\quad \epsilon\sim {R_p^{(7-p)/2}\over r_{UV}^{(5-p)/2}}\ , \qquad
d_{eff}={7-p\over 5-p}\ ,
\ee
using (\ref{Dpnullnorm}). The usual area law divergence for nonconformal 
theories on such a space, using (\ref{YMmetnonconf}), gives\
$S_{div} \sim N_{eff}(\epsilon) {V_{x^+} V_{y_i} \over \epsilon^{d-2}} 
= N (g_{YM}^2N)^{(d_{eff}-2)/2} \sqrt{Q} {V_{d-2}\over\epsilon^{d_{eff}-2}}$,\
of the form (\ref{nullEEnonconf}) in the bulk.

\section{Discussion}

We have described a null limit of entanglement entropy for theories
with a $g_{++}$ deformation (\ref{AdSnull}) arising in
gauge/string realizations of Lifshitz and hyperscaling violating
nonrelativistic holography, the strip subsystem stretched along a
null $x^+$-plane. This shows a milder leading divergence, perhaps
consistent with ultralocality in a field theory perspective, and
implied in the gauge/string realizations (\ref{AdSnull}) by
entanglement behaviour in hyperscaling violating backgrounds
\cite{Ogawa:2011bz,Huijse:2011ef,Dong:2012se}.

One can also consider null intervals, \ie\ 
spacelike strip subsystems with width $l$ along the $x^+$-direction,\
$\Delta x^+=-\Delta x^-={l\over\sqrt{2}} ,\ -\infty < y_i<\infty$, 
(although this has no natural interpretation after $x^+$-reduction). 
This gives
\bea\label{EEwidthx+}
\frac{\Delta x^+}{2} &=& \int^{r_*}_0
\frac{dr\ r^{d-1}}{\sqrt{A^2B^2+g_{++}r^{2d}-2Br^{2(d-1)}}} ,\quad 
\frac{\Delta x^-}{2}=\int^{r_*}_0
\frac{dr\ r^{d-1}(g_{++}r^2-B)}{\sqrt{A^2B^2+g_{++}r^{2d}-2Br^{2(d-1)}}} ,
\nonumber\\
S &=& {2R^{d-1}V_{d-2}\over 4G_{d+1}}
\int^{r_*}_{\epsilon}\frac{dr}{r^{d-1}}
\frac{AB}{\sqrt{A^2B^2+g_{++}r^{2d}-2Br^{2(d-1)}}} , 
\eea
from the entanglement area functional\
$S = {V_{d-2}R^{d-1}\over 2G_{d+1}}\int {dr\over r^{d-1}} \sqrt{1
- 2(\del_rx^+)(\del_rx^-) + g_{++}r^2 (\del_rx^+)^2}$.
For $AdS$ plane waves, this was studied in \cite{Narayan:2012ks}, and
a phase transition was observed, with no connected extremal surface 
for large $l$. The $\Delta x^+$ and $\Delta x^-$ integrals have 
widely different scaling. The boost\ $x^\pm\ra \lambda^{\pm 1} x^\pm$ 
scales up the $g_{++}$ term in the entanglement area functional and 
(\ref{EEwidthx+}) as before, eventually making $\Delta x^-$ positive 
(a spacelike subsystem requires $\Delta x^-<0$). The critical value 
occurs at $g_{++}r^2|_{r_c}\sim B$: for $AdS$ plane waves 
$r_c\sim Q^{-1/d}$.

Unlike (\ref{EEboostx}), (\ref{EEnull}), the null time $x^-$ slicing
does not arise as a highly boosted limit $\Delta x^+\sim l,\ \Delta
x^-\ra 0$, of entanglement (\ref{EEwidthx+}) for spacelike subsystems,
perhaps not surprising from the lower dimensional description.  The
boosted black 3-brane (which near extremality is the regulated $AdS_5$
plane wave) exhibits similar behaviour.  A null limit arises
\cite{Bousso:2014uxa} under a boost plus dilation, scaling down all
non-vacuum terms with the $g_{++}$ term being the most dominant: this
then becomes similar to the $AdS_5$ plane wave. The entanglement can
now be expanded and evaluated relative to the vacuum contribution
\cite{Bousso:2014uxa}:
$g_{++}\sim r^{d-2} T_{++}$ gives\ $\Delta S \sim \int g(x^+) 
\langle T_{++} \rangle$ with some function $g(x^+)$.

For the $AdS$-Lifshitz deformation, the entanglement is of the form
(\ref{EEwidthx+}), with $g_{++}=K^2$. As for $AdS$ plane waves, 
there is a phase transition here, with critical value\ 
$r_c\sim l_c\sim {1\over K}$.\  Under the boost plus dilation\ 
$x^+\ra x^+,\ x^-\ra\eta^2 x^-,\ r\ra\eta r$, with $\eta\ra 0$ 
as in \cite{Bousso:2014uxa}, the $g_{++}$ term does not scale down 
here, structurally similar to the $AdS_3$ plane wave (\ref{adspw}). 
The Lifshitz $g_{++}$ term is a non-normalizable deformation, sourced 
by \eg\ the lightlike dilaton/axion. 
Treating this as a small perturbation to $AdS$\ (with 
$\epsilon\ll {1\over K}$), we can expand (\ref{EEwidthx+}), as in 
\cite{Mukherjee:2014gia} for $AdS$ plane waves (and more generally 
\cite{Bhattacharya:2012mi,Allahbakhshi:2013rda,Wong:2013gua,
Blanco:2013joa,Lashkari:2013koa} for excited states), except this 
is not an excited state but a nontrivial CFT deformation. In this 
Lifshitz case,\  $r_*\sim l$, and\ \
$AB = r_*^{d-1}\sqrt{2B-K^2r_*^2}$\ \ from the turning point equation. 
Expanding about $B=1, K=0$ (as in $AdS$) gives
\be
S\sim {V_{d-2}R^{d-1}\over G_{d+1}} \int_\epsilon^{r_*} {dr\over r^{d-1}}
{ \sqrt{1-{K^2r_*^2/2B}} \over \sqrt{1-({r\over r_*})^{2(d-1)}}\ 
\sqrt{1-{K^2r_*^2 (1-(r/r_*)^{2d})\over 2B (1-({r\over r_*})^{2(d-1)})}}}\ 
\Rightarrow
\Delta S \sim {R^{d-1}\over 2 G_{d+1}} V_{d-2} K^2l^{4-d} {\cal M} ,
\ee
where the constant can be shown to be ${\cal M}>0$.\ 
There is no simple relation here between the $g_{++}$ deformation and 
the holographic stress tensor. It might be interesting to 
explore this.

\vspace{5mm}
\noindent {\small {\bf Acknowledgments:} I thank  A. Laddha, 
J. Maldacena and T. Takayanagi for helpful conversations, and N. Sircar 
and S. Trivedi for many discussions on entanglement and correlation 
matrices. I also thank the Organizers of the Strings 2014 conference, 
Princeton/IAS for hospitality while this work was in progress.}


\end{document}